\documentclass{webofc}
\usepackage[varg]{txfonts}   
\usepackage{orcidlink}

\begin{document}
\title{Quantum Computing for Nuclear Physics}
\author{\firstname{Martin J.} \lastname{Savage\,\orcidlink{0000-0001-6502-7106}}
\inst{1}\fnsep\thanks{\email{mjs5@uw.edu}}         }
\institute{InQubator for Quantum Simulation (IQuS), Department of Physics, University of Washington, Seattle, Washington, 98195, USA.}

\abstract{
Future quantum computers are anticipated to be able to perform simulations of 
quantum many-body systems and quantum field theories 
that lie  beyond the capabilities of classical computation.
This will lead to new insights and predictions  for systems 
ranging from  
dense non-equilibrium matter,
to low-energy nuclear structure and reactions, 
to high-energy collisions.
I present an overview of  digital quantum simulations in nuclear physics, 
with select examples relevant for studies of quark matter.
}
\maketitle
\section{Introduction}
\label{intro}
While remarkable progress continues to be made in understanding 
matter in extreme environments,
quantitatively extending far  beyond experiments and observations remains challenging with 
available analytic and computational techniques.
The limitations of classical computing in providing robust results for dynamics of quantum systems, 
or the nature of dense quantum systems, are well known, e.g. Ref.~\cite{NPexascale}. 
These limitations were anticipated in the pioneering works of Feynman~\cite{Feynman:1981tf}
and others, which identified quantum computation as a path forward.
Quantum computers are now a reality, and while evolving rapidly and growing in diversity and capabilities, 
are currently limited to modest-sized systems of noisy qubits and qudits, 
with relatively short quantum coherence times, i.e. we are in the Noisy Intermediate-Scale Quantum (NISQ) era~\cite{Preskill:2018jim}.  
The additional capabilities that quantum computation offers are the control over entanglement and over superposition,
which we are learning how to integrate into our computational toolbox and into analytic techniques.
With quantum advantages for specific computer-science problems, e.g. Ref.~\cite{Arute:2019zxq}, 
researchers are now vigorously pursuing quantum advantages for  scientific applications.
As our challenges in Standard Model physics
are intrinsically quantum mechanical, 
there is optimism that  they may furnish early demonstrations of quantum advantages for scientific applications.

Real-time time evolution
can be efficiently performed using an ideal quantum computer~\cite{Lloyd:1996}.
Therefore, if the relevant initial states can be prepared with sufficient precision,
future quantum computers hold the promise of simulating 
the time evolution of complex processes, such as hadronization and fragmentation, 
low-energy nuclear reactions, thermalization, 
coherent neutrino flavor evolution, and  the production of matter in the early universe, 
e.g., see Refs.~\cite{Bauer:2023qgm,Beck:2023xhh,Bauer:2022hpo}.
While  initial state preparation is not generally efficient at scale, even with quantum computers,
Nature is typically kind to us in this regard, with emergent symmetries, gaps and hierarchies, 
so that combinations of classical and quantum simulation are expected to provide initial states of 
sufficient fidelity for subsequent evolution in many instances.
Pioneering work by Jordan, Lee and Preskill (JLP)~\cite{Jordan:2012xnu} established 
a complete scaling analysis of lattice scalar field theory, from state preparation, 
through time evolution through particle detection.

While quantum computing eco-systems of the future may be 
able to facilitate end-to-end real-time simulations with a complete quantification of uncertainties,  
the near term will see them be used to study  aspects of such processes,
highlighting reaction pathways and the roles of entanglement and superposition.
While the former requires error-corrected quantum computers, the later can be accomplished 
to a certain extent with NISQ-era devices, particularly for processes 
where a result with a relatively large uncertainty, e.g. $10\%$ or even $100\%$, 
would constitute an improvement over what is possible with classical computation alone.

A growing array of quantum devices  are  being developed and co-designed for quantum simulations,
including superconducting qubits, trapped ions, neutral atoms, photonic systems, silicon quantum dots, and cavity-QED.
The different devices have different attributes -- different numbers of qubits or qudits, different coherence times, gate application times, gate fidelities, SPAM errors, and more -- such that different algorithms, software, workflows, and integrations into HPC environments, are being developed and benchmarked to simulate small model problems intended to advance our ability to simulate real-world systems in the future.  
Much of the device development is now being accomplished in technology companies,  
or in partnerships with national laboratories and universities, 
and a range of devices are available to the scientific community through cloud access.   
For instance, 
IonQ and Quantinuum are developing trapped-ion systems with more than 30 qubits and all-to-all qubit connectivity, 
and IBM is developing modular superconducting-qubit systems with hundreds approaching thousands of qubits with heavy-hexagonal connectivity. 
Nuclear physicists are working in these partnerships to explore paths forward for quantum simulations of  
strongly-interacting and correlated  systems important to Standard Model  physics research, 
currently by simulating systems of reduced complexity (e.g., reduced dimensionality)
but which share some key attributes.
These simulations
are advancing our way of thinking, our understanding of the role of superposition and entanglement,
and informing us about 
promising paths forward.
Directly including known properties of systems into quantum simulations, e.g., symmetries and confinement,
making them ``physics-aware'', is being recognized as being important
 to developing scalable circuits to perform simulations at scale, e.g. Ref.~\cite{Farrell:2023fgd}.

The distinction between analog and digital quantum simulation continues to blur.
In analog simulation, the target Hamiltonian/system is mapped to the native Hamiltonian of the device. 
In contrast digital quantum computers execute operations from a universal gate set.  
In principle, they can  simulate any quantum system with a given precision, 
and a challenge is to identify mappings of the Hamiltonian to unitary operations that 
scale in an efficient way to larger systems of experimental relevance.
In systems such as optical tweezer arrays of neutral atoms, one has both analog and digital capabilities (and HPC).

Mapping  quarks and gluons   
to available quantum computers is an example of current focus.
Formulated in the 1970's, the Kogut-Susskind lattice QCD Hamiltonian~\cite{Kogut:1974ag}  is defined in the chromo-electric basis, with links connecting states between neighboring lattice sites of a common irreducible representation of SU(3), nominally without bounds.  The chromo-magnetic interaction is included via plaquette operators, that can induce transitions between eigenstates of the chromo-electric operator.  
Of course, there are other mappings, and our community is benchmarking known implementations.
In moving toward  QCD,
quantum simulations are being performed of small and modest 1D spatial lattices for U(1), SU(2) and SU(3) systems, of differing particle content (for a recent review, see Ref.~\cite{Bauer:2022hpo}), with efforts into 2+1D and 3+1D.

\section{Lattice Gauge Theories and the Schwinger Model}
\label{LQCDSM}
\begin{figure}[!ht]
\centering
\includegraphics[width=4.7in]{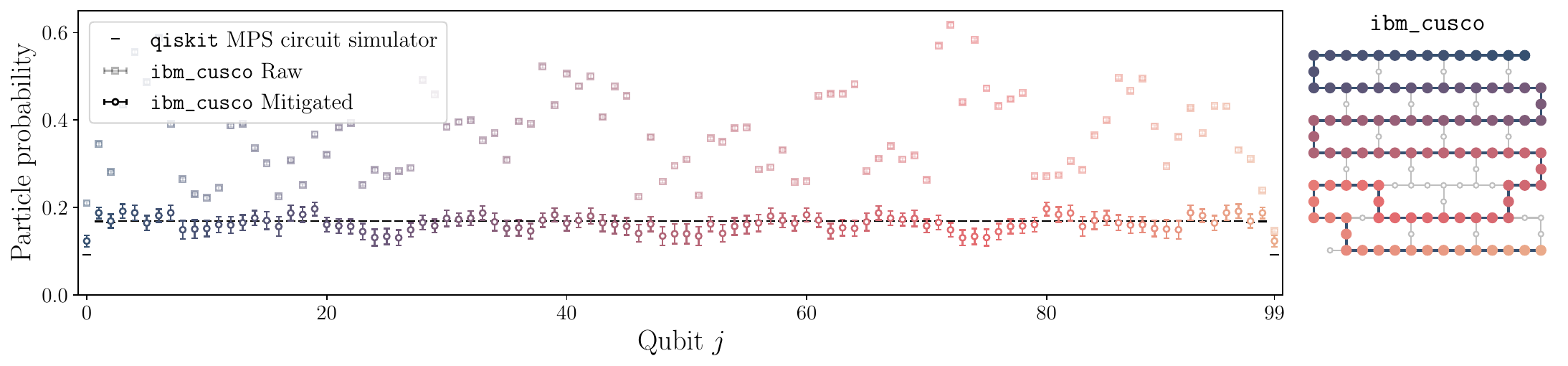}
\caption{The chiral condensate of the Schwinger model vacuum for a select set of parameters, prepared on 100 qubits of an 
IBM 127 qubit quantum computer~\cite{Farrell:2023fgd}.
}
\label{fig-1}       
\end{figure}

\noindent
For the reasons that the Schwinger model (1+1D quantum electrodynamics)
played an important role in developing QCD,  it is also playing an important role in developing quantum simulations of quantum field theories.
In 1+1D, there are no independent dynamical gauge 
degrees of freedom, the electric flux carried by a link is uniquely determined by Gauss's law, and different gauges can be chosen with different mappings to qubits -- in axial gauge, 
only the electrons and positrons are explicit and the effect of the gauge field is included via non-local ``Coulomb'' (charge-charge) interactions.  The theory is confining, so that electric charges are screened by the vacuum and its excitations are 
electron-positron composites.
The naive (vol)$^2$-scaling of contributions from the gauge-field Hamiltonian 
can be improved to  (vol)($\xi$)-scaling by considering confinement, 
where $\xi$ is the screening length, rendering it consistent with efficient scaling for quantum simulation.
Further,  compact circuit elements to prepare the vacuum and wavepackets can be determined classically (for a range of lattice spacings) and  scaled to arbitrarily-large quantum registers,
as shown in figure~\ref{fig-1}.
While the first quantum simulations of the Schwinger model in 2016 by an Innsbruck team 
used 4 qubits of a trapped-ion quantum computer~\cite{Martinez:2016yna},
simulations are now being performed with 100 qubits~\cite{Farrell:2023fgd}, 
with an eye toward particle collisions.
A number of different observables and phenomena have been explored in recent years, including string breaking and entanglement, energy-loss, particle scattering and collisions, e.g., Ref.~\cite{Florio:2023dke}.
Extensions to 1+1D SU(3) and SU(3) (QCD) have been accomplished for small systems, 
e.g., Ref~\cite{Ciavarella:2021nmj,Farrell:2022wyt}.   
Scaling these simulations to  larger systems is underway,
as are 
efforts to extend to Abelian and non-Abelian theories in
2+1D and 3+1D, where
both state preparation and time evolution of few-plaquette systems have been performed.
Important subsequent steps are to create wavepackets
and to understand how to detect composite particles produced in inelastic collisions.   
While particle detection has been developed for $\lambda\phi^4$ theory~\cite{Jordan:2012xnu}, 
it remains to be established for composite states.  
Current thinking is that including ``particle detectors'' into quantum simulations will resemble how actual experimental detectors work.   For example, the qubits of the simulated theory could be weakly coupled 
to other qubits programmed as a quantum sensor in a localized region of space, 
to measure, for instance, fluxes through a surface, or energy loss.

\section{Low-Energy Nuclear Physics}
\label{lenp}
\noindent
Progress has been made in simulating model systems on the path to low-energy nuclear structure and reactions, both using cloud-accessible quantum computers and also co-designed systems.  A hybrid technique has been used in describing the scattering of neutrons in which the spatial wavefunction is handled classically, while the spin degrees of freedom are distributed 
to the QPU, with an iterative feedback to update the spatial wavefunction~\cite{Turro:2023dhg}.   
There has been progress in using quantum algorithms to find ground state and low-lying excited states of light nuclei, for example $^6$Li, e.g., Ref.~\cite{Stetcu:2021cbj}. 
A recently developed algorithm, ADAPT-VQE~\cite{Tang:2019tpm} has been used in conjunction with a quantum emulator to demonstrate utility and limitations of straightforward approaches.  
A more comprehensive analysis has been performed for a range of encodings through the p-shell and sd-shell.
Entanglement structures and their rearrangement in nuclei have been investigated, and in models, 
such as the Lipkin-Meshkov-Glick  model, e.g., Ref.~\cite{Hengstenberg:2023ryt}.

\section{Thermalization}
\label{thermo}
\noindent
One of the important  areas that is starting to be addressed with quantum simulations is thermalization, 
a particular focus for QCD.
A rapidly growing number of studies in model systems are
now beginning to uncover fundamental aspects of thermalization of quantum systems, 
for example exploring the Eigenstate Thermalization Hypothesis (ETH)~\cite{Srednicki:1994ETH} and  beyond,
e.g., Ref.~\cite{Ebner:2023ixq}.
A typical simulation prepares an initial state and then ``quenches'' (corresponding to a rapid change in the corresponding Hamiltonian), 
and the subsequent time evolution of the system  leads to
expectation values of local operators that tend toward their thermal expectation values.
Such studies have already led to new understandings, insights, and directions of research, including into the role of entanglement.
The presence of scar states, typically distributed throughout the spectrum with anomalously low bipartite entanglement and which are only weakly coupled to the evolution Hamiltonian, 
delay thermalization after the quench.
It was thought that only confining theories displayed scar states, but recently it was demonstrated in a model theory that scar states are present in the de-confined phase, but  not  in the confined phase~\cite{Aramthottil:2022jvs}.

\section{Quantum and Thermal Phases}
\label{phase}
\begin{figure}[h]
\centering
\includegraphics[width=2.0in]{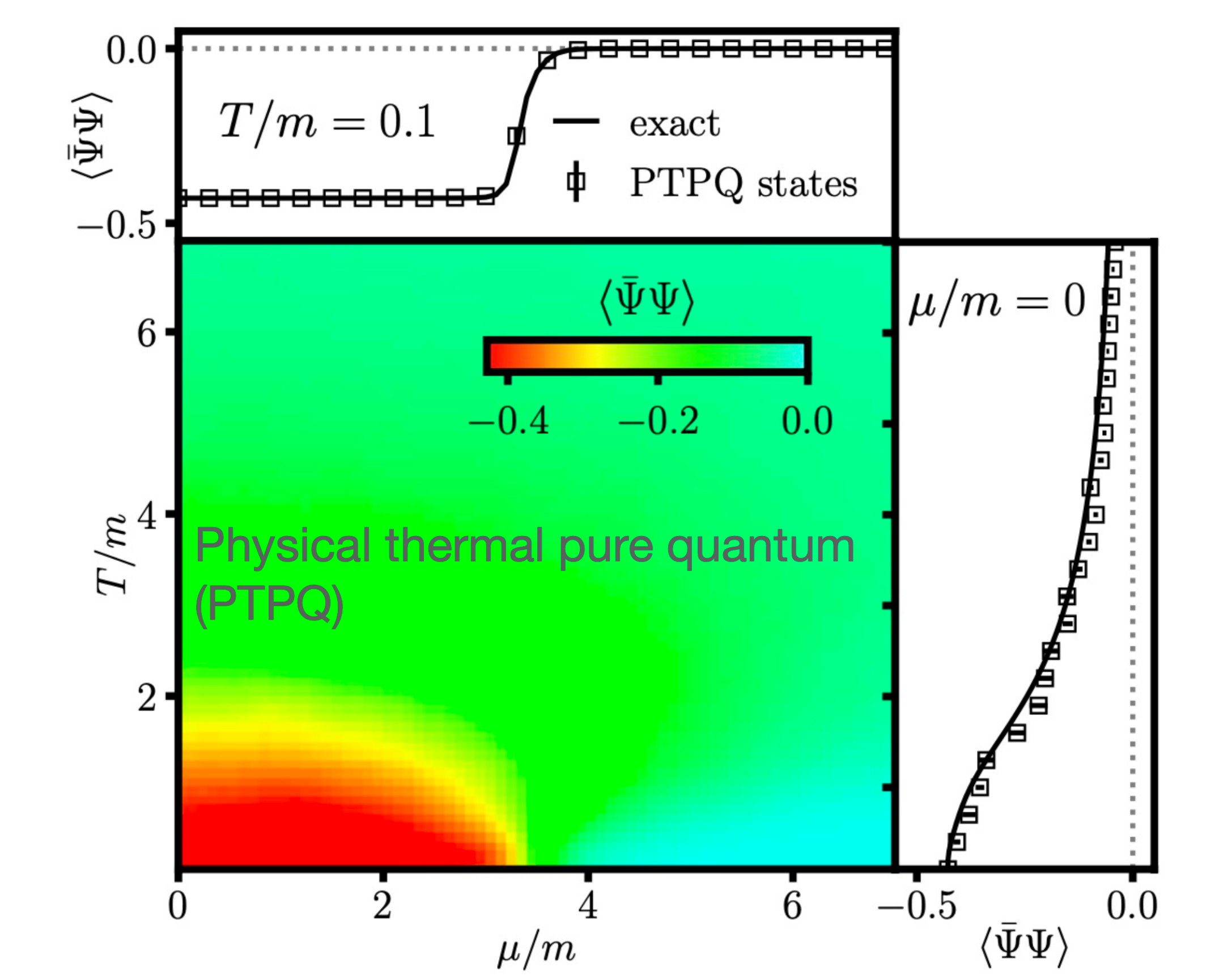}\ \ \ \ \ \ \ \ \ \ \ \ \ \ \ \ 
\includegraphics[width=2.1in]{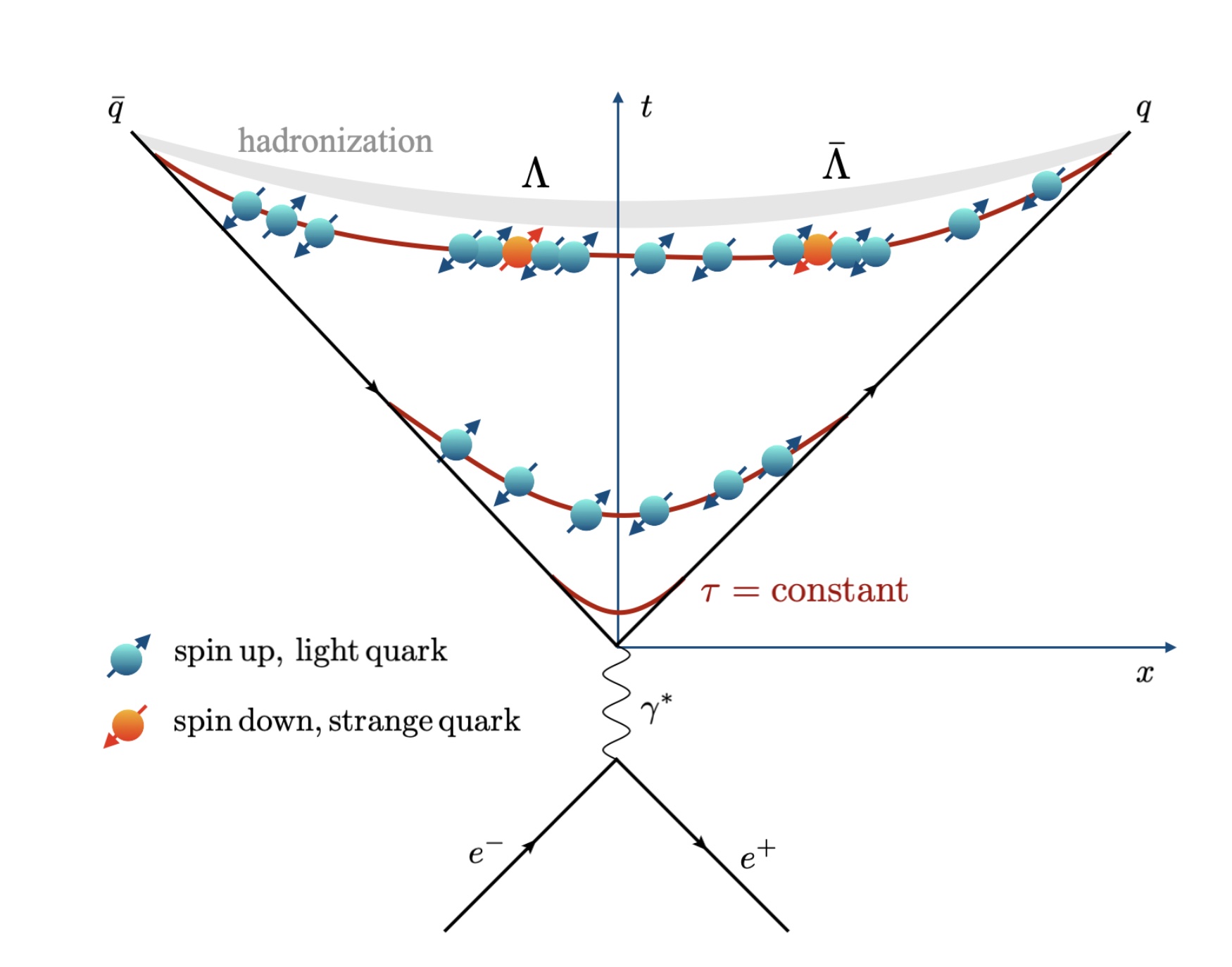}
\caption{
The left panel shows the phase diagram of a 1+1D $\mathbb{Z}_2$ lattice gauge theory computed using TPQ states~\cite{Davoudi:2022uzo}.
The right panel displays a cartoon of the spacetime evolution of a simplified Schwinger model simulation of  $\Lambda\Lambda$ spin correlations  produced in fragmentation~\cite{Barata:2023jgd}.
The left image is reproduced with permission from the authors~\cite{Davoudi:2022uzo} and under 
{\it Creative Commons Attribution 4.0 International license}.
The right image is reproduced with permissions from the authors~\cite{Barata:2023jgd}.
}
\label{fig-raju}       
\end{figure}

\noindent
There have been a number of recent demonstrations  of potential advantages of quantum simulations in model finite density systems.
One such demonstration was a simulation of the chiral condensate of the Nambu-Jona-Lasinio model with a (charge) chemical potential at finite temperature.  A quantum imaginary time evolution (QITE) algorithm was used to cool the system into its ground state for given temperature and chemical potential.  
A further recent development was a demonstration of the utility of thermal pure 
quantum states (TPQ states) in evaluating thermal averages in lattice gauge theories~\cite{Davoudi:2022uzo}.
The chiral condensate of the 1+1D $\mathbb{Z}_2$ lattice gauge theory was computed as a 
function of temperature and chemical potential, and was found to recover the exact result obtained using HPC, as shown in figure~\ref{fig-raju}.

\section{Entanglement}
\label{ent}
\noindent
Entanglement and quantum correlations in many-body systems can be probed in quantum simulations, 
providing new challenges for theorists.
1+1D models of quarks and string fragmentation have been simulated using quantum computers, with the goal of 
better understanding quantum correlations and potential violations of Bell's inequalities measurable in heavy-ion collisions, for instance in $\Lambda\Lambda$ spin correlations.
The dynamics of systems can be explored in real-time, 
as shown in figure~\ref{fig-raju}, 
and the amplitude, correlations and reaction pathway mapped out -- providing insights into the analogous process in QCD, see for example Ref.~\cite{Barata:2023jgd}.

\section{Neutrinos}
\label{neut}
\noindent
Neutrinos play an essential role in the evolution of supernova.
In the core, 
the density becomes sufficiently high during the collapse that the interactions between neutrinos dominate the 
flavor evolution.  
Ignoring off-forward scattering, the effective Hamiltonian  maps to an all-to-all connected
spin model.  
The two-flavor system maps straightforwardly onto a trapped-ion quantum computer, and the all-to-all connectivity allows for relatively straightforward quantum circuits to time evolve prepared initial states.
A number of simulations of (two-flavor) systems with up to 20 neutrinos have been performed on quantum computers and simulators, e.g., Ref.~\cite{Amitrano:2022yyn}.  
Starting from tensor-product states, the evolution of  entanglement has been explored,
including multi-neutrino entanglement~\cite{Illa:2022zgu}.
Further, dynamical quantum phase transitions have been identified in these system, separating fast and slow modes, 
e.g., Ref.~\cite{Roggero:2021fyo}.

\section{Discussion}
\label{disc}
\noindent
We are developing understandings, techniques, and the necessary collaborations, 
to simulate quantum field theories and quantum many-body systems using quantum computers 
by identifying and simulating less complex problems that share key features of target systems. 
It is  plausible 
that a quantum advantage could be demonstrated for nuclear physics processes 
using NISQ-era quantum computers, but significant challenges remain.

\section{Acknowledgements}
\label{disc}
\noindent
I thank my colleagues for all of the insights and wisdom they have imparted.
Page limitations preclude providing many of the appropriate references - my sincerest apologies.
This work was supported by U.S. Department of Energy, Office of Science, Office of Nuclear Physics, 
InQubator for Quantum Simulation (IQuS)\footnote{\url{https://iqus.uw.edu/}}
under Award Number DOE (NP) Award DE-SC0020970 
via the program on Quantum Horizons: QIS Research and Innovation for 
Nuclear Science\footnote{\url{https://science.osti.gov/np/Research/Quantum-Information-Science}},
and by the Department of Physics
and the College A+S
at the UW.


\end{document}